\documentclass[preprint,12pt,3p]{elsarticle}
\usepackage{graphicx,pslatex,float,color,array,amssymb,amsmath,euscript}
\usepackage{subcaption}
\usepackage{lipsum}
\usepackage{siunitx}
\usepackage{xcolor}
\usepackage{graphicx}
\usepackage{lipsum} 
\usepackage{tabularx}
\usepackage{dblfloatfix}
\usepackage[position=b]{subcaption}
\usepackage{caption}

\definecolor{Blue}{rgb}{0.3,0.3,0.9}
\definecolor{Std}{rgb}{0.,0.,0.}
\def\rev#1{\color{Std}{#1}\color{Std}}
\usepackage{caption}
\journal{Superalloys 2024}

\begin{document}
\begin{frontmatter}
\title{Influence of the coating brittleness on the thermomechanical fatigue behavior of a $\beta$-NiAl coated R125 Ni-based superalloy}
\author[add1]{Capucine Billard}
\author[add2]{Damien Texier}
\author[add1]{Matthieu Rambaudon}
\author[add1]{Jean-Christophe Teissedre}
\author[add3]{Noureddine Bourhila}
\author[add3]{Dimitri Marquie}
\author[add3]{Lionel Marcin}
\author[add3]{Hugo Singer}
\author[add1,cor]{Vincent Maurel}

\cortext[cor]{Corresponding author. E-mail address: vincent.maurel@minesparis.psl.eu (V. Maurel)}
\address[add1]{MINES Paris, PSL University, MAT - Centre des Matériaux, CNRS UMR 7633, BP 87 91003 Evry, France}
\address[add2]{Institut Clement Ader (ICA) - UMR CNRS 5312, Université de Toulouse, CNRS, INSA, UPS, Mines Albi, ISAE-SUPAERO, Campus Jarlard, 81013 Albi Cedex 09, France}
\address[add3]{SAFRAN Aircraft Engines, Site de Evry-Corbeil, Rue Henri-Auguste Desbruères, BP 81, 91003 Evry Cedex, France}

        \begin{abstract}
                 The brittleness of an aluminide diffusion coating protecting a Ren\'{e}~125 Ni-based polycrystalline superalloy was investigated \rev{over } a wide range of temperatures in its as-received and thermally aged form. Isothermal and thermal \rev{cycled } aging were performed on the coated system at a \rev{maximum } temperature of \SI{1100}{\celsius}. Microstructure evolutions and damage initiation within the coating were characterized. Interrupted tensile tests and thermomechanical fatigue tests were conducted to document critical stress-strain conditions leading to the coating cracking and lifetime for \rev{the case of } thermo-mechanical fatigue loading. Advanced digital image correlation and acoustic emission techniques were used to detect coating cracking. Isothermal oxidation or cyclic oxidation \rev{led } to improved strain-to-failure due to metallurgical  evolutions and also \rev{longer } fatigue life under thermomechanical fatigue conditions.

        \end{abstract}

\end{frontmatter}
	\section{Introduction}

    To withstand high temperature and oxidative environment, Ni-based superalloys are typically coated with an alumina-forming or chromia-forming material, depending on the temperature window/environmental conditions. Nickel aluminide (NiAl) coatings on  Ni-based superalloys are a common type of protective coating for high-temperature turbine blades~\cite{Mevrel:1987}. They are designed to act as a local aluminum reservoir to form a protective, temperature-stable, and adherent alumina scale. However, NiAl coatings can have detrimental effects on the mechanical response of the overall system due to \rev{their } brittle behavior up to 650-\SI{700}{\celsius}~\cite{Mevrel:1987,Pan:2003,Alam:2013,Texier:2020a}. This brittle-to-ductile transition makes \rev{them } susceptible to through-thickness cracking at low temperatures during a service cycle~\cite{Eskner2003,alam2010evaluation}. In fact, such external cracks can lead to premature failure of the coated blade and \rev{by propagating } into the superalloy substrate under cyclic thermomechanical loading~\cite{Evans:2009}. The life gain from improved oxidation performance can be largely impaired by the brittle behavior of the coating. Therefore, the detection and prediction of crack initiation under experimental conditions close to those in service is essential for estimating the lifetime of the entire system. 
    For high temperature loading, typical of thermomechanical fatigue (TMF) loading, early damage is critical due to the way stress can occur at a low temperature range, especially when considering out-of-phase  thermomechanical fatigue (OP-TMF) conditions~\cite{Remy:2000val,Caron:2011sx}, Figure \ref{fig:TMFload}. 

    Exposure to high temperature and oxidizing environment induces a strong evolution of the NiAl coating microstructure~; initially consisting of a homogeneous layer of typical $\beta$-NiAl-rich phase, the outer layer of the coating is prone to $\gamma'$ precipitation due to high temperature oxidation and Al consumption to form the external alumina-scale but also the interdiffusion with the Ni-base superalloy substrate~\cite{Mevrel:1987,Angenete:2004,maurel2022coated}. This is the result of the interdiffusion flux of Al for both the outward diffusion flux from the coating to the surface to form the thermally grown oxide (TGO), mainly Al$_2$O$_3$, and the inward Al diffusion flux from the coating to the substrate together with the flux of Ni from the substrate to the coating. In addition, $\gamma'$ formation is accelerated when thermal cycling is considered compared to isothermal oxidation, \rev{this effect being mainly driven by cyclic oxide spallation } \cite{Sallot:2015,Maurel:2016rumpling}. It has been observed that, the room temperature strain-to-failure is increased with this formation of $\gamma'$ phase, $e.g.$ for (Ni,Pt)Al~\cite{Esin:2016a}. However, no clear conclusion has been proposed on the effect of aging on the brittle-to-ductile transition temperature (DBTT). Along with this issue, the mechanism of the observed increase in strain-to-failure is still an open question. \rev{On the one hand the $\gamma'$ phase fraction increases more rapidly for thermal cycling than for isothermal oxidation for a given time spent at maximum temperature, with different localisation and morphology of $\gamma'$ precipitates. On the other hand, } the surface roughness  \rev{increases more rapidly considering  thermal cycling as compared to isothermal oxidation } by the so-called rumpling effect~\cite{Tolpygo:2008}. \rev{Both phase transformation and roughness increase modify the ductility of the coating because the $\gamma'$ phase is more ductile than the $\beta$ phase and because the roughness increases the local stress value by stress concentration effect, limiting the apparent ductility of the coating}. The above two questions are important to provide a clear vision of the benefits of coating pre-aging, as well as an assessment of the risk of failure in the context of thermal cycling and TMF loading. 
    
\begin{figure}[h] %
    \centering
    \includegraphics[width=.5\textwidth]{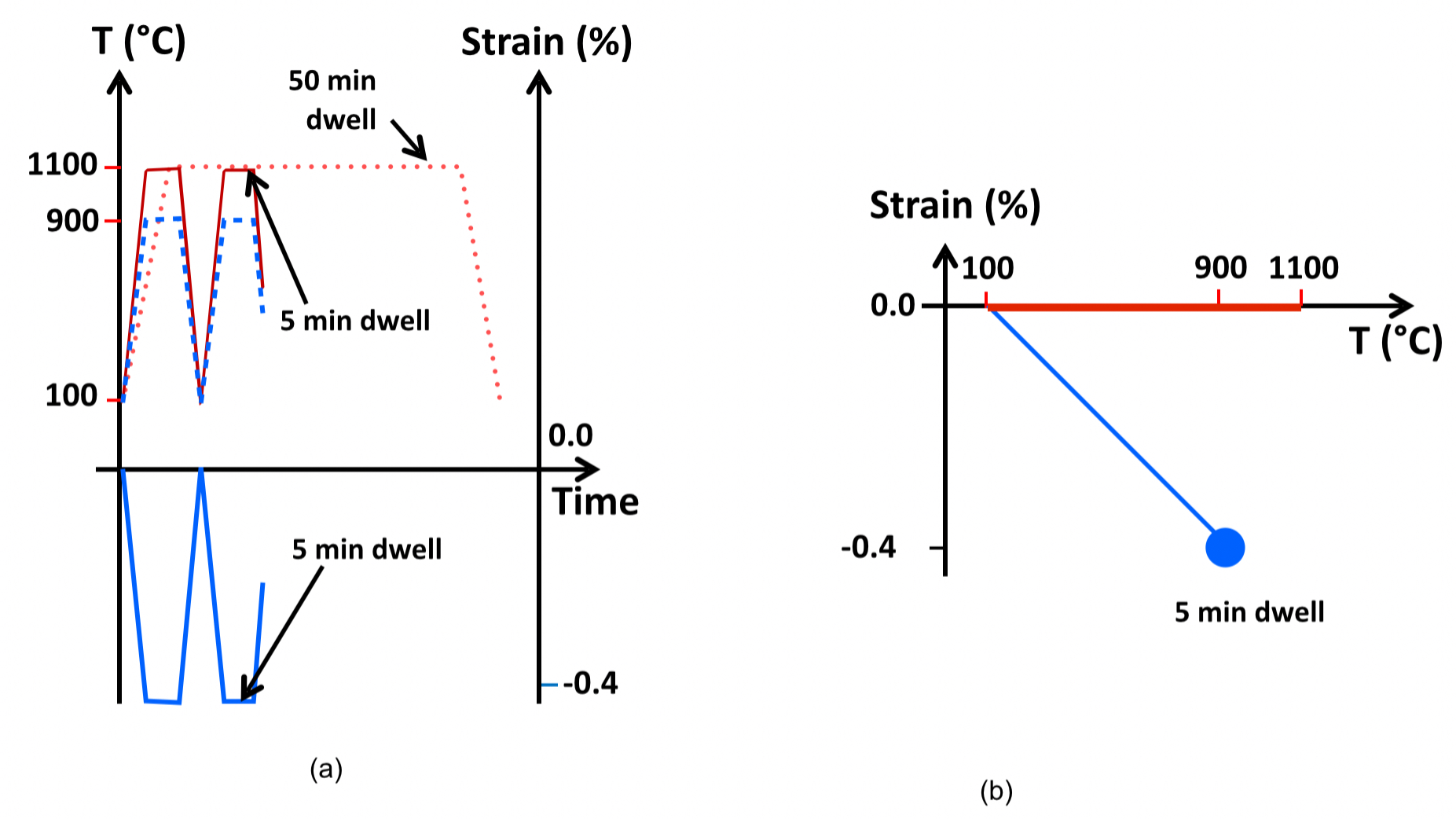}
    \caption{Applied (a)~thermal cycling with different hold times and strain controlled TMF testing (b)~resulting strain to temperature OP-TMF cycle (in blue) compared to pure thermal cycle (in red)}
    \label{fig:load} %
\end{figure}
    
    The purpose of this study is to analyze the effect of both isothermal oxidation and thermal cycling on the critical strain-to-failure as a function of the  temperature. This investigation has been conducted on a typical \rev{low activity and high temperature } NiAl coating deposited onto a \rev{cast } Ni-based polycrystalline superalloy (Ren\'{e}~125). Special attention is given to the strain localization and damage mechanisms using mesoscale and microscale digital image correlation (DIC) and acoustic emission (AE) techniques on as-received and pre-aged specimens. Coated systems were subjected to either incremental tensile or TMF testing to \rev{generate } critical macroscopic stress/strain conditions leading to the coating cracking. On this basis, a discussion of the observations made with typical OP-TMF loading is proposed.

\begin{figure*}[h] %
    \centering\includegraphics[width=0.95\textwidth]{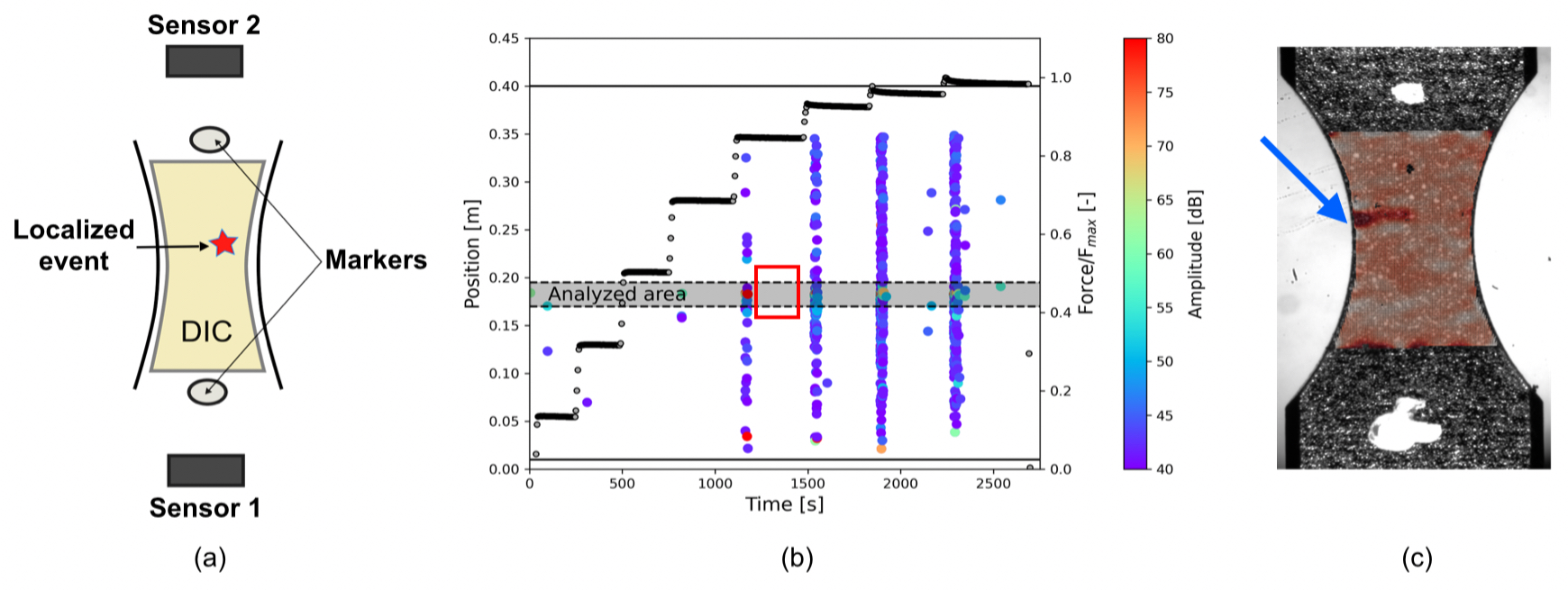}  
    \caption{Mesoscale set-up (a)~sketch of markers and AE sensors (b)~evolution of force and acoustic events as a function of time (c)~DIC analysis of strain component in the loading direction}
    \label{fig:mesosketch} %
\end{figure*}

	\section{Experimental methods}
	\subsection{Material and specimens}
    The studied material consists of a NiAl coating deposited onto a Ren\'{e}~125 Ni-based polycrystalline superalloy by \rev{low activity } vapor phase  aluminization (APVS: Aluminizing Phase Vapor Snecma) followed by a heat treatment at \SI{870}{\celsius} for 16~hours.  The coating is an intermetallic compound made up of two characteristic layers: (i)~the outer layer composed mainly of the $\beta$-NiAl phase and (ii)~the interdiffusion zone (IDZ) composed of the  $\gamma'$-Ni$_{3}$Al phase and topologically compact phase precipitates (TCP). A hyperstoichiometric layer forms in a shallow region below the outer surface~\cite{oskay2017evolution}. Due to its higher aluminum concentration, this layer is particularly brittle and will be prone to cracking under thermal and/or mechanical stress. 
    \begin{table}[h]
        \centering
            \caption{Composition in coating (coat.) and R125 substrate (sub.) measured by EDS averaged on the thickness of the coating (\%at)}
    \begin{tabularx}{\linewidth}{XXX XXX XXX X}
             
               &Ni&Al&Cr&Ti&Ta&W&Co&Mo&Hf \\\hline
             Coat.&bal.&39.9&4.5&1.0&0.6&0.5&7.7&0.4&0.6\\
             Sub.&bal.&11.8&8.2&3.5&1.1&1.6&9.0&0.9&0.7
        \end{tabularx}
        \label{tab:R125compo}
    \end{table}

    \subsection{Isothermal oxidation, thermal cycling and thermo-mechanical fatigue conditions}
    Thermal aging treatments were performed at \SI{1100}{\celsius} in air under both isothermal and cyclic conditions. The thermal cycles included heating from 100 to \SI{1100}{\celsius}, then holding for 5~min or 50~min, and finally cooling to room temperature. 
    
    A schematic illustration of the thermal cycle is depicted in Figure~\ref{fig:load}(a).
    
In order to get a clearer picture of the global behavior of NiAl coated R125, out-of-phase \rev{strain controlled } thermomechanical fatigue (OP-TMF) cycles were performed. This OP-TMF loading combines a thermal cycle with a 5~min dwell at the maximum temperature set to \SI{900}{\celsius}, with a strain cycle from 0 to -0.4~\%, with a dwell time at -0.4~\% corresponding to the dwell at the maximum temperature, Figures~\ref{fig:load}(a) and~(b).

\begin{figure*}[!ht] %
    \centering
    \includegraphics[width=0.95\textwidth]{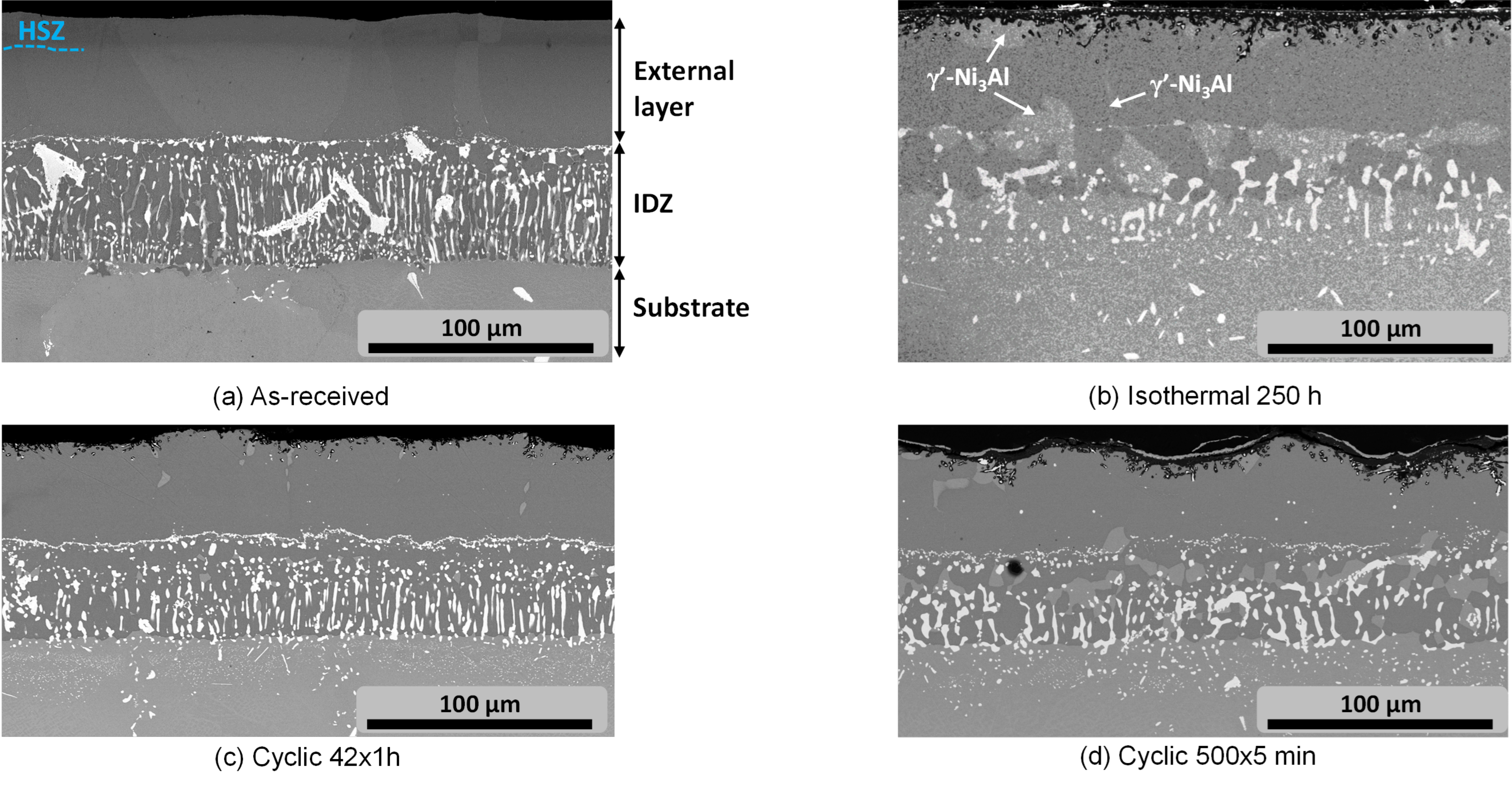}
            \caption{Microstructure  for different aging conditions observed in BSE mode (a)~AR state, (b)~isothermal oxidation for 250~h at \SI{1100}{\celsius} (c)~thermal cycling 1~h dwell at \SI{1100}{\celsius} and N=42 and (d)~thermal cycling 5~min dwell at \SI{1100}{\celsius} and N=500}
    \label{fig:evolmicro} %
\end{figure*}
 
    \subsection{Tensile testing and mesoscale analysis}
    The ductility of the whole system and crack initiation behavior of the external coating were studied on the as-received and aged materials by comparison at different temperatures. To study crack initiation, tensile tests were performed at least up to the brittle-to-ductile transition temperature.
    Tensile specimens were designed as dog-bone specimens with a rectangular cross-section and  cross-sectional variation to drive the maximum stress and strain localization to the center of the specimen, where the cross-section is a square of 2x\SI{2}{mm^2}. This specimen geometry is particularly useful for concentrating and tracking deformation/damage events using image acquisition during testing~\cite{Esin:2016}. A lamp furnace and K-type thermocouple allow the specimen temperature to be controlled up to \SI{900}{\celsius}. Incremental tensile loading is then prescribed by displacement steps until cracks are detected in the specimen gauge. The critical local strain required for crack initiation in the coating is obtained with accurate \textit{in-situ} observations of each side of the specimen using a Keyence VHX 1000 microscope and a CCD camera for further digital image correlation (DIC) with VIC-2D software. \rev{In the AR state, the natural roughness induces enough contrast for DIC. After aging,  to improve the quality of DIC, speckle pattern was obtained by white paint spots on a black paint used as a background}. These observations are also combined with two acoustic emission (AE) sensors to detect and localize events during loading. AE greatly improves the resolution of the method in terms of detecting the first crack to appear (Figure~\ref{fig:mesosketch}~(a) and~(b)). \rev{The sketch in Figure~\ref{fig:mesosketch}~(a) presents the association of AE sensors, surrounding the gauge length of the specimen, in this latter markers are used to track global displacement and speckle is used to monitor strain field through DIC. As an example, the strain field obtained by DIC for the as-received state is shown in Figure~\ref{fig:mesosketch}~(c). The Figure~\ref{fig:mesosketch}~(b) combines the prescribed displacement with black markers, the colored markers represent the intensity of AE, analyzed area being highlighted by a grey area: the chosen threshold for detection consisted in AE signal level higher than 70 dB when associated to strain localization observed by DIC}.

    \subsection{Micro-testing and OHR-DIC at room temperature}
    Interrupted tensile tests were conducted at room temperature on flat micro-tensile specimens in the as-received and aged states. \textit{Ex-situ} optical high resolution-digital image correlation techniques (OHR-DIC) were conducted using an OLYMPUS LEXT OLS5100 laser scanning confocal microscope (LSCM) to document onset of cracking. LSCM allowed for focused image acquisition regardless of the surface topography. A gold speckle pattern similar to Ref.~\cite{Rouwane2023} was applied to track surface kinematics after deformation. Large mosaic of images were acquired before and after deformation (13~$\times$~7~images with an overlap of 20~$\%$ and a pixel size of 205~nm). Non-periodic optical distortions were corrected similarly to Ref.~\cite{Rouwane2023}.  DIC calculations were performed using fast optical flow algorithms~\cite{Kroeger:2016}.

\section{Results}
    \subsection{Impact of aging on microstructure evolution and damage}
    As a first intent, the evolution of microstructure from as-received (AR) state to different agings was herein described (see Fig.~\ref{fig:evolmicro}).

        \subsubsection{As-received microstructure}
        The microstructure in the AR condition is typical of a diffusion coating,  Figure~\ref{fig:evolmicro}(a), showing (i)~an external layer composed mainly of $\beta$-NiAl (dark gray), with the darkest contrast showing the hyper stoichiometric zone (HSZ in blue)  closer to the coating surface, (ii)~an interdiffusion zone (IDZ), with a matrix composed of $\beta$-NiAl and a rather high density of precipitates \rev{presumed to be }  TCP (light particles), and then (iii)~the superalloy with a few TCP precipitates, eutectic particles and the well known $\gamma$/$\gamma'$ microstructure.

        The thickness of the outer layer is about 58~$\pm$~10~µm (deviation indicating differences observed from sample to sample, rather than within a given cross-section), with \rev{the } hyperstoichiometric layer limited to \SI{10}{\micro\meter}. The surface roughness of the outer layer is initially slightly marked, with peaks in the core of $\beta$-NiAl grains and valleys in the vicinity of grain boundaries, for grains reaching \SI{100}{\micro\meter}. The roughness of the outer layer/IDZ interface is similar to the former. This interface is marked by a bright line of precipitates with no residual grit particles. 

        The total thickness of the IDZ is approximately 54~$\pm$~3~µm. The IDZ is composed of two distinct regions: the region closer to the outer coating is composed of equiaxed precipitates, a region of 9~$\pm$~1~µm thickness, and the region closer to the substrate is composed of elongated precipitates, a region of 48~$\pm$~2~µm thickness. The direction of precipitate elongation corresponds to the direction of diffusion flux. There are also a few bright areas of $\gamma'$ matrix in this region. From this BSE contrast it can be concluded that small grains, in the range of 10~µm, of $\beta$ matrix are present in the vicinity of the outer coating/IDZ interface, while larger grains are observed in the outer layer. 

        It should be noted that this AR layered structure is consistent with the entire test series. Only TMF tests were performed on thicker coatings, for which the microstructure \rev{evolution with high temperature exposure is very similar to observations made above}.

        \subsubsection{Isothermal oxidation}
        After isothermal oxidation, the surface of the coating shows complex oxidation patterns and fine precipitation in the vicinity of initial $\beta$ grain boundaries (Figure \ref{fig:evolmicro}(b)). Bright particles and light gray particles are observed, the latter being associated with $\gamma'$ phases. The same contrast is observed closer to the coating surface and near or at the outer coating/IDZ interface. The surface roughness remained similar to the one of the AR state. Within the IDZ, a large amount of $\gamma'$ phase is observed with equiaxed grains close to the outer coating/IDZ interface and a sort of diffusion front in the region of elongated TCPs. The substrate exhibits elongated TCPs that are very numerous compared to the AR state.
        
        \subsubsection{Thermal cycling}
        For 1~h dwell at high temperature, the surface roughness slightly increased compared to the AR state, which could be related to the so-called rumpling phenomenon, Figure \ref{fig:evolmicro}(c) and (d). In addition, oxide pegs with sizes similar to the isothermal state are observed, although the accumulated time at high temperature is lower for the observed cyclic case. Some $\gamma'$ precipitates are observed close to the surface, and at the $\beta$ grain boundaries, but the bright particles are not observed in the outer layer. 

        Within the IDZ, the evolution of the matrix is similar to the observations made for the isothermal case, also the TCPs in the substrate are less numerous. This point is consistent with the fact that the diffusion time for 42~cycles of 1~hour is very low compared to 250~hours for isothermal oxidation. This also confirms that the $\gamma'$ transformation in the outer layer is the result of a strong coupling with thermal cycling, \rev{by both recrystallisation of grains and oxide spallation, as suggested in~\cite{Sallot:2015}}. 

        For \rev{tests with a } 5-min. dwell at high temperature, the outer surface becomes rougher than in any other case, while the time accumulated at high temperature is relatively low (about 42~h of exposure at \SI{1100}{\celsius}), Figure \ref{fig:evolmicro}(d).
 
         Meanwhile, the phase transformation from $\beta$ to $\gamma'$ is more pronounced than for the 1-h dwell condition, and this is observed both in the outer coating and in the IDZ. Oxidation also appears to be more damaging than for 1 h dwell, as some pegs could be associated with initial cracking in the outer part of the coating.

        In conclusion, diffusion controlled the evolution from $\beta$ to $\gamma'$ along an evolution of the matrix of the IDZ, also a diffusion front controlled by the cumulative time at maximum temperature. The $\gamma'$ phase also precipitated at $\beta$ grain boundaries in the outer layer. Thermal cycling promotes surface undulation, the rumpling phenomenon, and damage associated with oxide pegs.

    \subsection{Strain-to-failure evolution with temperature and aging}
	\subsubsection{The mesoscale DIC and acoustic emission}
        The coating brittleness of AR and aged coated systems was evaluated as a function of the temperature using interrupted tensile tests paired with mesoscale DIC and acoustic emission (AE). The strain-to-failure corresponds to the macroscopic strain at which the first crack is observed within the external coating at a given temperature. 

\begin{figure}[!ht] %
    \centering
    \includegraphics[width=\columnwidth]{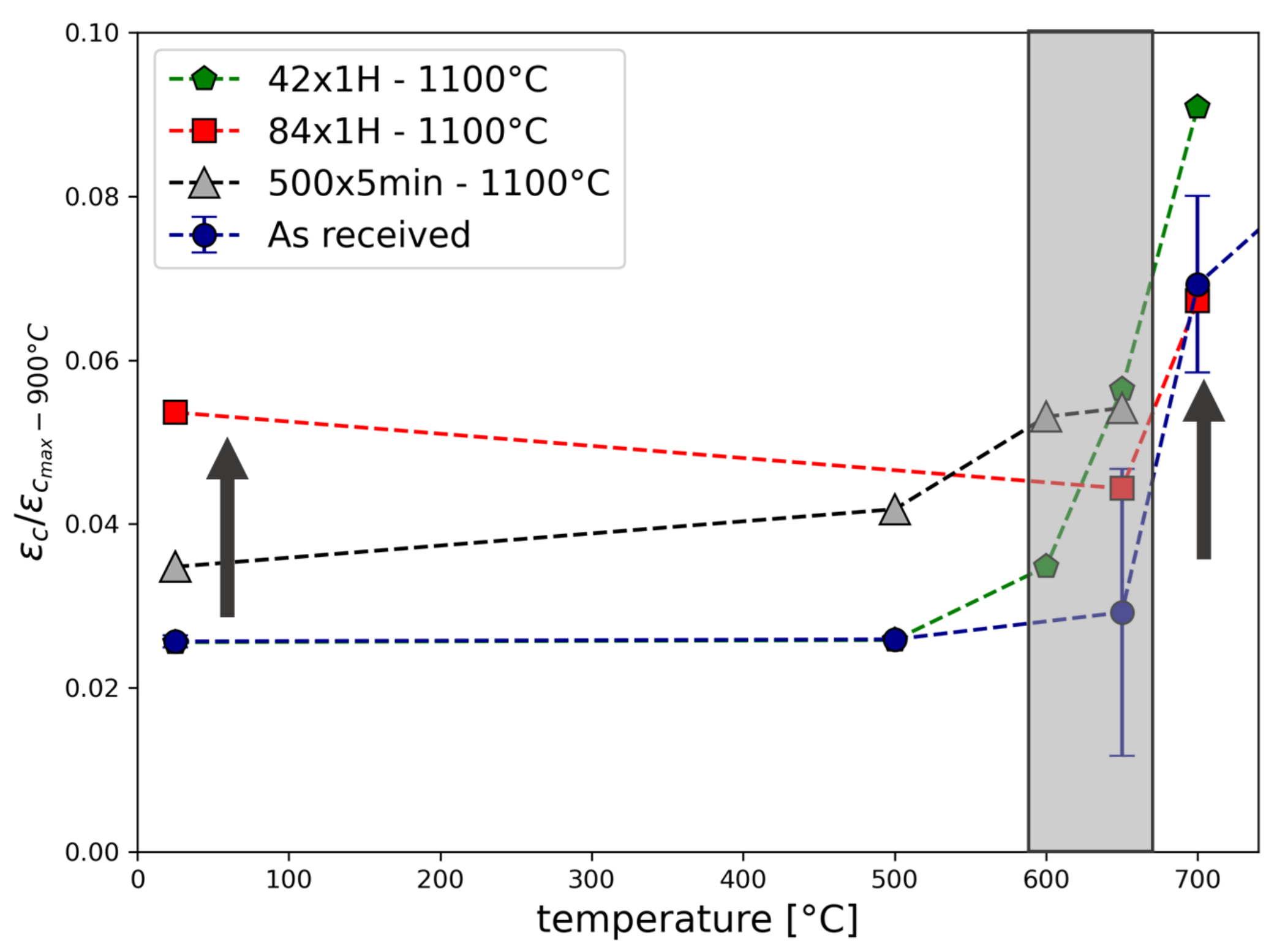}
    \caption{Strain-to-failure evolution with temperature and aging}
    \label{fig:defcrit} %
\end{figure}

        The as-received coating was found to be particularly brittle from room temperature to \SI{700}{\celsius}, Figure \ref{fig:defcrit}. The associated cracks were observed to reach the whole width of the specimen in a single step of loading. This is evidenced by strain localization patterns observed by DIC, see \rev{the } arrow Figure \ref{fig:mesosketch}(c). For temperature equal or higher than \SI{700}{\celsius}, the ductility increase is obvious, for test carried out at \SI{900}{\celsius} no crack was observed, this maximum strain being used to rationalize the experimental strain values for each test ($\varepsilon_{c\max}$ \SI{900}{\celsius}). 
 
        After thermal aging, a low number of thermal cycles with long dwell time at maximum temperature does not impact the room temperature ductility (see green curve corresponding to 42~cycles for 1-h dwell at \SI{1100}{\celsius} in Figure~\ref{fig:defcrit}). However, a slight gain in ductility is observed for this condition of aging for temperature of \SI{600}{\celsius} and higher. 

\begin{figure*}[!ht] %
    \centering
    \includegraphics[width=.95\textwidth]{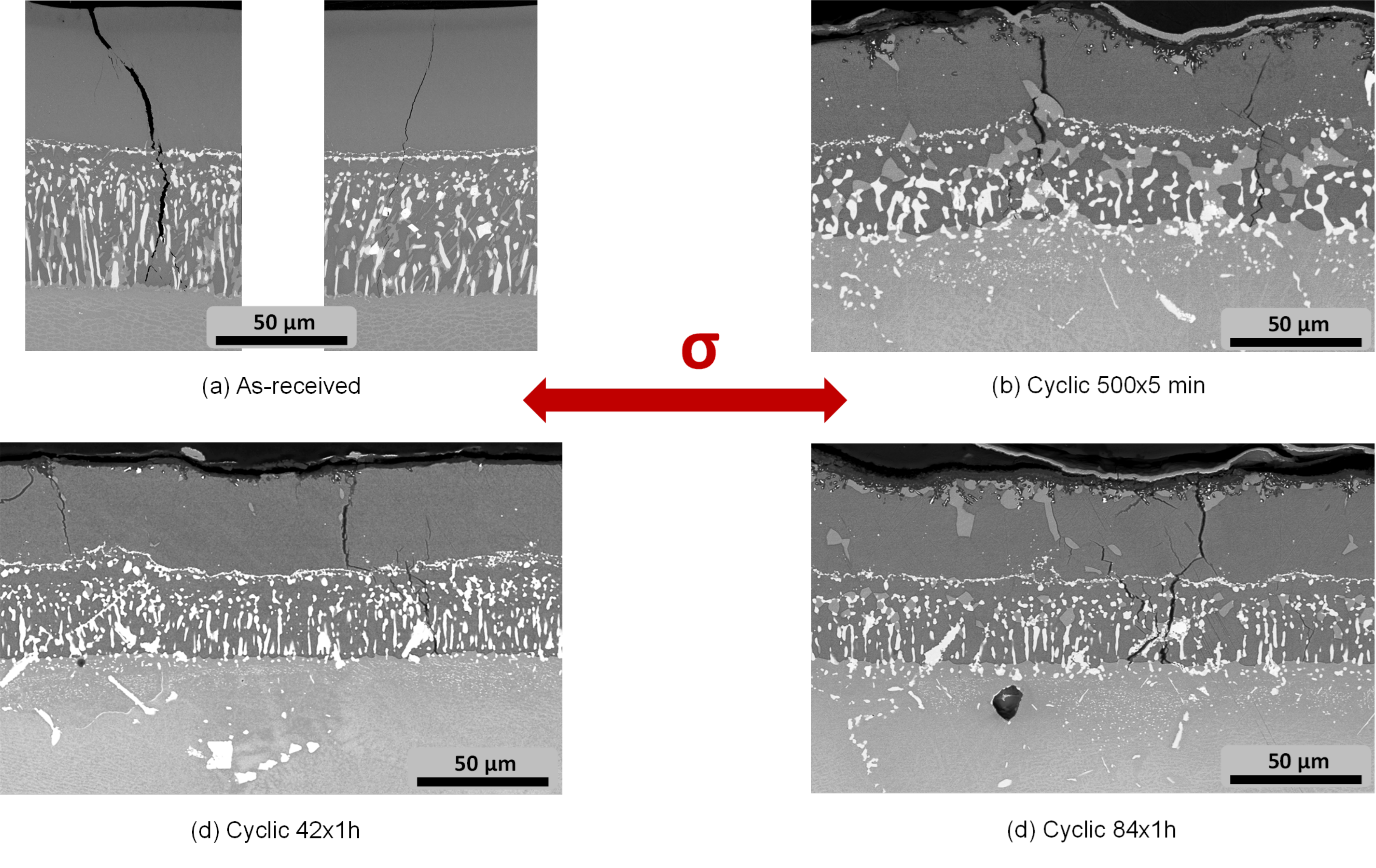}
    \caption{Microstructure and crack after \rev{room temperature } tensile test for different aging conditions (a)~AR state, (b)~thermal cycling of 5-min dwell at \SI{1100}{\celsius} and N=500, (c)~thermal cycling 1-h dwell at \SI{1100}{\celsius} and N=42 and (d)~thermal cycling 1-h dwell at \SI{1100}{\celsius} and N=84}
    \label{fig:crack} %
\end{figure*}

        For longer high temperature exposure at maximum temperature or shorter dwell time, the ductility at room temperature has been clearly increased, as illustrated in Figure~\ref{fig:defcrit} (gray and red curves, corresponding to 500~cycles for 5-min dwell and 84~cycles for 1-h dwell at \SI{1100}{\celsius}, respectively). It is noteworthy that the time spent at maximum temperature is of about 42-h for 500 cycles of 5-minutes, that is to say that the dwell time impacts the ductility of the considered material. 

        Thermal aging alters the microstructure of the outer layer of the coating, resulting in a significant increase in strain to failure in room temperature tensile tests. In order to gain insight into the crack evolution within the coating and its relationship to the microstructure, $i.e.$, virgin or aged, $post mortem$ sections were observed using SEM after discontinuous tensile tests at room temperature. These micrographs allow a direct comparison between the as-received and thermally aged systems. It is important to note that the cracks are modified by the microstructure. 
        
        For AR state, the cracks easily join the substrate to IDZ interface despite some local \rev{deviations } (\rev{two cracks are shown in } Figure~\ref{fig:crack}(a)). In addition to very limited ductility, the through "coating plus IDZ" thickness cracking are very detrimental for the integrity of the system, directly impacting the failure of the substrate. For 42~cycles of 1~h, crack opening is more limited, and the crack path is clearly more tortuous  (Figure~\ref{fig:crack}(c)). Furthermore, crack deviations are observed at the external coating/IDZ interface. This could impede the crack growth even though in the given observation, the crack in the IDZ reaches again the IDZ/substrate interface. For 84~cycles of 1~h, \rev{the } crack path seems to be connected to $\beta/\gamma'$ interfaces in the external coating (Figure~\ref{fig:crack}(d)). Again this should limit the crack extension even though the density of crack is higher for 84~cycles of 1~h aging than for 42~cycles of 1~h aging. For 500 cycles of 5 min, crack paths are also driven by $\beta/\gamma'$ interfaces, but could also be observed in a pure $\beta$ region, associated  with \rev{both } $\beta$ grain boundaries \rev{and not } (Figure~\ref{fig:crack}(b)). However, cracks are tortuous for both cases. The brightest particles in the external coating do not seem to impact cracking. The role of surface roughness associated with rumpling on the coating cracking is not obvious from these observations. This point will be addressed in \rev{a follow-on paper}. 
        
        To summarize these results, the ductility can be increased after significant aging inducing microstructure evolution. 

\begin{figure*}[!b] %
    \centering
    \includegraphics[width=0.95\textwidth]{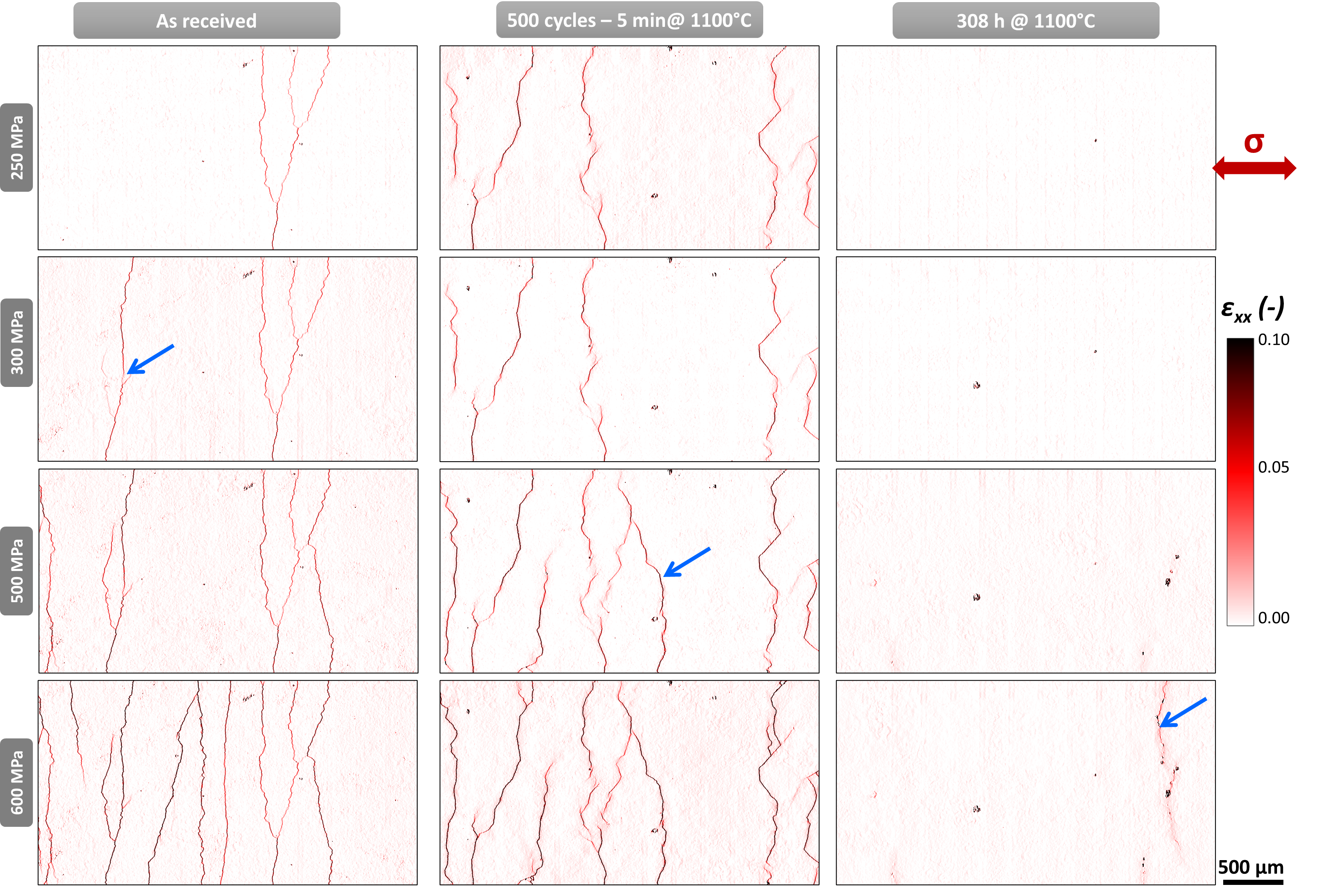}
    \caption{Strain maps along the macroscopic tensile direction for the different loading conditions, for the different aging conditions.}
    \label{fig:DICmicroAll} %
\end{figure*}
 
    \subsubsection{Strain localization and cracking at the microstructure scale using optical high resolution - digital image correlation (OHR-DIC)}

     OHR-DIC was carried out on rough original surfaces using LSCM in AR and aged states, using a specimen of uniform section within the gauge length. The strain observations correspond to the total width of the gauge length,  Figure~\ref{fig:DICmicroAll}.

    In the AR state, cracks developed across the gauge width for macroscopic stresses as low as 250~MPa (related to a strain of 0.13~$\%$), as shown in Figure~\ref{fig:DICmicroAll}. New cracks continuously developed with the macroscopic deformation, with the first significant crack event highlighted with blue arrows.

    For the 5-min dwell thermal cycling condition, short cracks were found for stress as low as 250~MPa (related to strain of 0.13~$\%$). The cracks did not propagate across the full gauge width as compared to the AR condition. The crack pattern associated with this short dwell time is denser than that observed for the AR state, but also exhibits crack branches that are more limited in length and more tortuous. Further crack growth was observed as the deformation increased up to 500~MPa. %

    For the 308~h isothermal oxidation condition, coating cracking was only observed at 600~MPa. Therefore, the aging \rev{not only } changed the ductility of the coating system but also its fracture mechanism.

    \subsection{Thermomechanical fatigue behavior}
\begin{figure*}[!ht] %
    \centering
    \includegraphics[width=.95\textwidth]{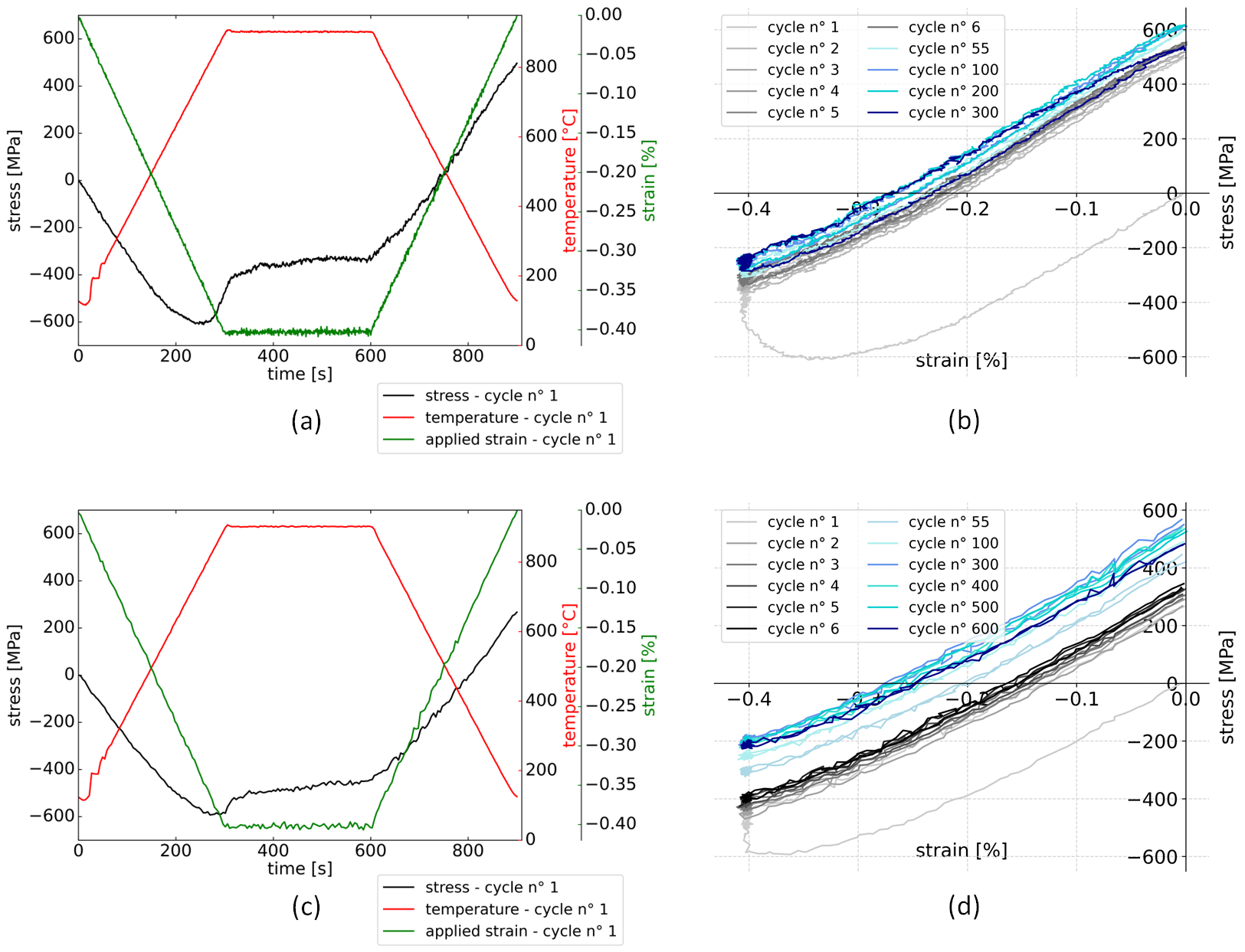} 
    \caption{TMF temperature-strain-stress evolutions \rev{for both as received state (AR) and  pre-aging, corresponding to 42~cycles of 1~h dwell at \SI{1100}{\celsius}} (a)~function of time for AR specimen, (b)~stress-strain function of the number of cycles for AR specimen, (c)~function of time for pre-aged specimen, (d)~stress-strain function of the number of cycles for pre-aged specimen.}
    \label{fig:TMFload} %
\end{figure*}

    Thermomechanical tests were conducted to evaluate the fatigue life of the coated system for out-of-phase  thermomechanical fatigue (OP-TMF) conditions, and the impact of prior high temperature aging, implying modification of the coating, interdiffusion and substrate, on the TMF lifetime. The chosen OP-TMF cycles induced a complex behavior considering the stress-strain hysteresis plot. For the first cycle, the compressive stress monotonically increased up to a deformation of $\varepsilon$=-0.3\%, followed by a stress plateau and a final compressive stress decrease due to the very high temperature (Figure~\ref{fig:TMFload}(a)). At the maximum compressive strain, the stress relaxation is about \SI{100}{MPa}. Then, when the strain goes back to zero along with the temperature decrease, the stress relaxation induces large tensile stresses at the lowest temperature (500~MPa at the first cycle). It is worth noting that the stress relaxation decreased in amplitude with further cycling (Figure~\ref{fig:TMFload}(b)). However, both stress relaxation and \rev{progressive evolution of stress-strain loops by cyclic kinematic hardening, that could be associated to the so-called }  Bauschinger effect~\cite{Lemaitre:1990}, increased the maximum tensile stress (610~MPa at the $\#$100~cycle). This point is critical considering the observed low ductility of the coating for temperatures below \SI{600}{\celsius}, as reported in the previous sections using tensile testing. 
        
 The impact of pre-aging on the durability within such TMF loading will be discussed in  a \rev{follow-on paper}.

\section{Discussion}
    The present paper focuses on the impact of aging conditions on microstructural evolutions and changes in strain-to-failure behavior of a coated Ni-based superalloy system (here, $\beta$-NiAl coated R125). The ultimate motivation here was to impose quite severe OP-TMF cycles to evaluate the gain in lifetime with thermal aging of the system. 
    
    \subsection{Impact of microstructure evolution of the coating on strain-to-failure}

    Thermal aging, in an isothermal or cyclic way, lead to microstructural evolutions of the coated system due to the combined effect of the oxidation and diffusion with the Ni-based superalloy substrate (See Figure~\ref{fig:evolmicro}). The main evolutions consisted in $\beta$ to $\gamma'$ transformation in both the external coating and in the IDZ matrix. In the external coating, $\gamma'$ precipitation is mainly localized at grain boundaries, promoted by thermal cycling. These microstructural evolutions are controlled by the diffusion front related to the accumulated time spent at maximum temperature. In addition, thermal cycling promotes surface undulation, the so-called rumpling phenomenon, and damage associated with oxide pegs, \rev{that are potential sites for crack initiation}. Repetitive short-dwell thermal cycling was found \rev{to be } particularly detrimental for the coated systems due to the occurrence of short cracks within the external coating system. Such microstructural evolutions are in good agreement with the literature on aluminide coating systems~\cite{Monceau:2001,Sallot:2015,EvansHE:2011}. The major difference from previous studies for (Ni,Pt)Al systems is the large amount of TCP phases within the outer coating and within the substrate \rev{after } long-term exposure~\cite{Sallot:2015}.

    The effect of these microstructural changes on the brittleness of the coated system was investigated from room temperature to \SI{900}{\celsius} using different approaches. Incremental tensile tests paired with mesoscale DIC and AE aimed at detecting the initiation of the first crack as a function of the macroscopic stress/strain level for different temperatures (Figure~\ref{fig:defcrit}). This approach aimed at documenting the effect of thermal aging on both the strain-to-failure \rev{capability } of the coated system and its ductile-to-brittle transition temperature (DBTT). $\beta$ to $\gamma'$ transformation in the outer layer lead to an increase in ductility of the coated system, the coating being particularly brittle when the coating is purely composed of $\beta$ phase and in the presence of hyperstoichiometric layer (i.e., without significant $\beta$ to $\gamma'$ transformation). Thus, the aging should be long enough to promote homogenization of the composition, and a significant volume fraction of $\gamma'$ precipitates to gain ductility. However, it is also obvious that the time spent at high temperature modifies the behavior of the coating but also of the superalloy due to modification of the $\gamma$/$\gamma'$ microstructure. The DBTT was not significantly affected by the thermal aging despite the microstructural evolutions.

    As far as the brittleness of the coating is concerned, cross-sectional observations (Figure~\ref{fig:crack}) and microscale DIC techniques aimed at evaluating differences in crack path depending on the thermal aging condition (Figure~\ref{fig:DICmicroAll}). It is worth mentioning that long-term isothermal oxidation can significantly improve the cracking resistance of the coating system (here 308~h at \SI{1100}{\celsius}). This is in line with in-depth study of NiAl evolution with aging as developed for example in~\cite{oskay2017evolution}. Indeed, the coated system experienced cracking at 250~MPa in its as-received state compared to 600~MPa for the long-term isothermal state. In addition to the onset of crack development, the crack pattern also \rev{sheds light on the } coating brittleness. Through coating plus IDZ cracking and fully transverse cracking are signs of high brittleness. Shorter cracks and crack arrests/bifurcations within the IDZ testify to the influence of microstructural change on \rev{improving the } ductility of the coated system. It is worth noting that the surface crack pattern observed with OHR-DIC (Figure \ref{fig:DICmicroAll}) is fully consistent with the thermal cycling crack observed through the cross-section (Figure \ref{fig:crack}): the cracks are mostly limited by either $\beta/\gamma'$ interfaces or $\beta$ grain boundaries. 
    
    This result is particularly important, showing that prior aging can be a solution to improve the overall ductility of such coated systems under the DBTT. Another message is the fact that the ductility of the coated system will present a significant gain during service if $\beta$ to $\gamma'$ transformation occurs. As aforementioned, the severity of the thermal cycling using short-dwell cycling can introduce damage and early cracking of the coating. This damage feature should explain that the ductility reaches a sort of asymptotic regime at 5~minutes dwell, because the benefit of the $\gamma'$ transformation is \rev{balanced } by the introduction of damage at these short cycles.
    
    \subsection{Impact of aging on thermomechanical fatigue failure}
  
    To clarify the way aging could impact the behavior of the studied system, TMF loading were tested after prior aging at high temperature. 
    Pre-aged specimens under thermal cycling conditions were also tested in similar out-of-phase  thermomechanical fatigue (OP-TMF) conditions. Since short cycles are detrimental in terms of coating damage, we preferred to use 42 cycles of 1~h dwell at \SI{1100}{\celsius} as pre-aging.
    
    Stress-strain hysteresis was plotted in Figure~\ref{fig:TMFload}(c) for the first cycle. The stress-strain-temperature evolution is very close to the AR condition, an increase of the compressive stress is found as a function of the temperature increase. However, less pronounced stress-relaxation during the strain dwell led to moderate tensile stresses at the lowest temperature (200~MPa lower than AR condition). The Bauschinger effect is yet more pronounced between the $\#$6 and $\#$55 cycles for the aged specimen, but the maximum tensile stress after $\#$300~cycles (and even $\#$600~cycles) is lower than the maximum tensile stress experienced during the OP-TMF test in the AR condition. Interestingly, the aged specimen had a two times greater lifetime under the same OP-TMF conditions compared to the AR specimen.

As a first \rev{observation}, it is obvious that the mechanical behavior has been modified by this pre-aging: this should be related to the fact that the pre-aging should have affected the microstructure of both the coating and the substrate, \rev{since } the maximum temperature during aging was  \SI{1100}{\celsius} \rev{which exceeds } the  \SI{870}{\celsius} \rev{post-coat heat treatment}. Further investigation is underway to determine the extent to which pre-aging has modified the superalloy microstructure (in particular the $\gamma-\gamma'$ ratio and the precipitations of TCPs), the coating microstructure (in particular the phase transformation) and morphology (in particular its roughness).

Last but not least, the life extension associated to pre-aging demonstrates that optimal heat treatment should consider the coating system in the sense that a given diffusion coating is impacted by the choice of the superalloy. This offers new guidelines to obtain a full optimization of the coated superalloy due to the drastic impact of the composition of this latter \rev{on this effect }~\cite{leng2020effect}.

\section{Concluding remarks}
Several methods were used to determine the risk of crack initiation in a NiAl diffusion coating protecting a typical Ni-based polycrystalline superalloy, here R125. First, a mesoscopic DIC measurement coupled with AE was used to easily detect the first crack initiation. On this basis, both room temperature ductility and DBTT were observed to increase with prior aging. For a given thermal cycling condition, the strain-to-failure is increased with the progressive transformation from $\beta$-NiAl to $\gamma'$-Ni$_3$Al. This important result confirms previous results obtained for the (Ni,Pt)Al coating of a single crystal superalloy. However, a detrimental effect of short-dwell thermal cycling was also observed. Therefore, \rev{the optimal thermal conditions for aging are still not fully clear}. Isothermal aging is not promoted here since the transformation from $\beta$-NiAl to $\gamma'$-Ni$_3$Al in this case significantly affects the outer coating/IDZ interface. Cross-sectional observations and OHR-DIC on the rough coating surface aimed at \rev{discerning } differences in the damage mechanisms \rev{evident}. Indeed, the as-received state is \rev{highly brittle and is associated with } full transverse cracking across the specimen gauge width, and through thickness coating plus IDZ cracking. Aging led to crack propagation barriers within the IDZ with several crack bifurcations observed, but also short cracks in the rough coating surface. Finally, OP-TMF with a dwell time in compression induces large stress relaxation at high temperature and large tensile stress at low temperature leading to potential early coating cracking. Prior thermal aging has improved the TMF life of the coated system, which could be seen as a new guideline for optimizing the heat treatment method for diffusion coated superalloys. 

 \bibliographystyle{unsrt}

 \bibliography{refs.bib,gradbib.bib}
		
	\end{document}